\documentclass[10pt]{enoc2011}
\usepackage{epsfig,amsmath,amsfonts}

\title{Cosmic ray propagators \vspace{0.2cm}\\in the fractional differential model of bounded anomalous diffusion\vspace{0.4cm}}

\author{Vladimir V. Uchaikin, Renat T.
Sibatov}

\address{Department of Theoretical Physics, Ulyanovsk State University, Ulyanovsk, Russia}

\abstract{Fractional differential approach to cosmic ray physics
problems is discussed. A short review in this field is given, some
results are represented, analyzed and criticized. A new model
called the bounded anomalous diffusion model is offered. Its
equation includes the fractional material derivative which allows
to take into account the finite speed of cosmic rays particles.}

\begin{document}

\pagestyle{myheadings}

%%%%%%%%%%%%%%%%%%%%%%%%%%%%%%%%%%%%%%%%%%%%%%%%%%%%%%%%%%%%
\section{Introduction}

For a long time, cosmic rays (CR) propagation through the Galaxy
have been described in frame of the normal diffusion theory, based
on the equation
$$
\left[\frac{\partial }{\partial t}-C\triangle
\right]\psi(\mathbf{r},t)=\delta(t)\delta(\mathbf{r}),
$$
where $C$ is the diffusion coefficient independent on frequency.
This equation is derived under the assumption that fluctuations of
interstellar magnetic fields, giving particle trajectories a
specific (Brownian) character, are characterized by definite space
sizes and look more homogeneous at large scales.

Rising attention to fractional calculus and its applications
haven't passed CR physics as well (see \cite{1}). An operator only
symbolically different from the fractional derivative with respect
to time, appeared (being unrecognized) in \cite{2}. The same
derivative we meet in the equations derived by \cite{3,4} (see
also the reviewing part of recent work \cite{5}). Replacing the
first time-derivative in the diffusion equation by its fractional
analog reflects slowing-down of CR diffusion caused by magnetic
traps (see \cite{6}). This kind of diffusion process was called
\textit{subdiffusion}.

At the same time, the turbulent (fractal) character of the
interstellar magnetic field (see \cite{7,8}) accelerates the
diffusion. The turbulent diffusion (\textit{superdiffusion}) was
described by \cite{9} by means of diffusion equation with
fractional (2/3) power of Laplacian. \cite{10} and \cite{11}
combined both of these regimes (super- and subdiffusions) to one
anomalous diffusion process, described by the space-time
fractional equation
$$
\left[_0\textsf{D}_t^\beta+C(-\triangle)^{\alpha/2}\right]
\psi(\mathbf{r},t)=\frac{t^{-\beta}}{\Gamma(1-\beta)}\delta(\mathbf{r}),\eqno(1)
$$
Its solution depends on two parameters: fractional orders of the
partial derivatives with respect to the coordinates ($\alpha\in(0,
2]$) and time ($\beta\in(0, 1]$). The characteristic features of
this solution are the presence of heavy power-law tails (for
$\alpha < 2$) and the law of diffusion packet spreading
$t^{\beta/\alpha}$. The family of distributions was named
\textit{fractional stable distributions}  by \cite{12}, and
investigated later in detail.

First application of fractional diffusion model to CR physics was
connected with energy spectrum problem (Lagutin and Uchaikin,
\cite{13,14,15}). The fractional diffusion model was used because
interstellar magnetic field heterogeneities were testified to take
large-scaled (fractal) character~(see \cite{Kul:94}). The
supernova remains analysis shows the presence in this region of
gas components with different physical parameters
($T_{e}\sim5\div10^6$~K, $n_e\sim0.1\div10^3$~m$^{-3}$), what can
be the sequence of extreme heterogeneity of interstellar medium.
These facts and other data concerning the heterogeneities of
matter density $\rho$ and magnetic field intensity
$H\propto\rho^q$, $q\sim1/3\div1/2$~(\cite{7}) within the length
range 100-150~pc, giving rise to uncertainties of diffusion model
to be applicable to CR transfer description, stimulated the use of
the superdiffusion model, based on the fractional Laplace
operator:
$$
\left[\frac{\partial }{\partial
t}+C(E)(-\triangle)^{\alpha/2}\right]
\Phi(\mathbf{r},t,E)=S(\mathbf{r},t,E).\eqno(2)
$$
This model seemed to be able to explain the experimentally
observed "knee" in the energy spectrum, i.~e. increasing of the
exponent $\eta$ in power representation of the $E^{-\eta}$
spectrum while passing from the region $\sim 10^2$~Gev/nucleon
into that of $\sim10^5$~Gev/nucleon. The explanation was connected
with the presence of power kind asymptotics of the solution at
small and large distances from a point source. Moreover, this was
in accordance with the self-similarity hypotheses which led Monin
to the same equation in the turbulent diffusion problem.

Later, Lagutin et al. published a number of works in which the
distributions $\psi(\mathbf{r},t)$ obtained by \cite{10,11} were
compared with experimental data by choosing the parameters. In the
process of those investigations, the parameters were changed from
initial values $\alpha = 1.7,\ \beta =1\div 0.8$ used in our first
works in 2001 to $\alpha = 0.3$ and $\beta = 0.8$ in 2004 (see
\cite{16}, p.~13); thus, the ratio $\beta/\alpha$ increased from
0.47 to 2.67. The former values were not unreasonable, whereas the
latter values seem to be absurd. Indeed, in this case, the cloud
of CRs instantaneously emitted by a point source spreads according
to the law
$$
R_c\propto t^{\beta/\alpha} = t^{2.67},
$$
whereas the limit speed of the particles is the light speed $c$
and the diffusion packet cannot spread faster then $R_c=ct$.

The cause of such conclusion is that Eqs.~(1) and (2) describe
processes of instantaneous jumps separated by random periods of
immobility state. This contradiction to real physical process was
first overlooked because of the absent of a clear interpretation
of fractional derivatives.

\section{On fractional derivatives and fractals}

The pecularity of fractional derivatives is their separation from
the usual differential and the related notion of increment. The
representation that the increment $\Delta y$ is for some reasons
proportional to ($(\Delta x)^\nu$)í does not concern the
Riemann-Liouville derivative and its three-dimensional Riesz
generalization. In this case, one has to consider a special form
of the increment, the difference of the fractional ($\nu$-th)
order:
$$
\Delta^\nu y=\left(1-e^{-\Delta x\frac{d}{dx}}\right)^{\nu}y(x)=
\sum\limits_{k=0}^\infty(-1)^k{\nu\choose k}y(x-k\Delta x).
$$

However, this difference cannot be interpreted in an explicit way,
when $\nu$ is not an integer. For this reason, the derivation of
fractional equations usually begins with integral relations, which
are then subjected to the Fourier-Laplace transformations
$\mathbf{r}\mapsto \mathbf{k},\ t\mapsto \lambda$ and asymptotic
expansions. As a result, the products
$|\mathbf{k}|^{\alpha}f(\mathbf{k},\lambda),\ \alpha\in (0,2],$
and $\lambda^\beta f(\mathbf{k},\lambda),\ \beta\in(0,1],$ are
treated as the Fourier transform of the fractional Laplacian and
the Laplace transform of the Riemann-Liouville fractional
derivative of the orders $\alpha$ and $\beta$, respectively,
$$
|\mathbf{k}|^{\alpha}f(\mathbf{k},\cdot)=\int
e^{i\mathbf{k}\mathbf{r}}\triangle^{\alpha/2}f(\mathbf{r},\cdot)d\mathbf{r},\qquad
\quad \lambda^\beta f(\cdot,\lambda)=\int\limits_0^\infty
e^{-\lambda t}\ _0\textsf{D}_t^\beta f(\cdot, t)dt.
$$

The very expressions $|\mathbf{k}|^{\alpha}f(\mathbf{k},\cdot)$
and $\lambda^\beta f(\cdot,\lambda)$ appear as asymptotic
expressions (for $\mathbf{k}\to 0, \lambda\to 0)$), and inverse
transformations using Tauberian theorems lead to the power-law (in
$\mathbf{r}$ and $t$) functions in the asymptotic limit of large
values of arguments. Such distributions characterize, in
particular, fractal structures and processes whose main property
is self-similarity. For example, if a sequence of random points is
located on the real axis so that the distances $R_j$ between the
neighboring points are identically distributed and independent and
the distribution of points is self similar on average (i.e.,
$\langle N (x) \rangle=\langle N (1)\rangle x^\nu,\ \nu\in(0,1]
$), the equation for the distribution density of $R_j$ includes
the fractional derivative of the order $\nu$. The resulting
distribution of random points has all of the attributes of a
stochastic fractal, including intermittence. As $\nu$ increases,
these attributes weaken and the equation with $\nu = 1$ is the
first-order equation whose exponential solution provides the model
of an inhomogeneous Poisson ensemble, which has a homogeneous form
at large scales. This is the relation of fractal structures with a
fractional derivative.

Since the most important property of space plasma is turbulence,
manifesting self-similar structures of a power-law (fractal) type,
the application of the fractional derivative technique to
diffusion in this medium seems to be appropriate in this case.

Application of the fractional differential equation (1) to
anomalous diffusion is based on the continuous time random walk
model introduced by \cite{20}. In this model, the walk of a
particle is represented by a sequence of instantaneous jumps with
random lengths at random times between which the particle is at
rest. The lengths of the jumps and intervals during which the
particle is at rest (in ``a trap'') are independent of each other.
To illustrate the relation of the walk scheme with the real
transport of CRs in the galactic magnetic field, let us divide
space into cubic cells and specify each particle inside the $i$-th
cell by the radius vector $\mathbf{r}_i$ of the center of this
cell (Fig. 1). At a random time $T$ after entering this cell, the
particle passes to one of the six neighboring cells and the
corresponding vector jumps to the center of this new cell at the
time of the intersection of the interface between the old and new
cells. After a random time $T'$, the particle passes to the next
neighboring cell and the vector again instantaneously jumps, etc.
If these cells were identical in the properties, the walk of the
specifying vector would be discretized Brownian motion, which is
transferred to ordinary diffusion with an increase in the scale.
In case of a regular medium all cells are identical and free
motion regime is absent.

\begin{figure}[tb]
\centering
\includegraphics[width=0.8\textwidth]{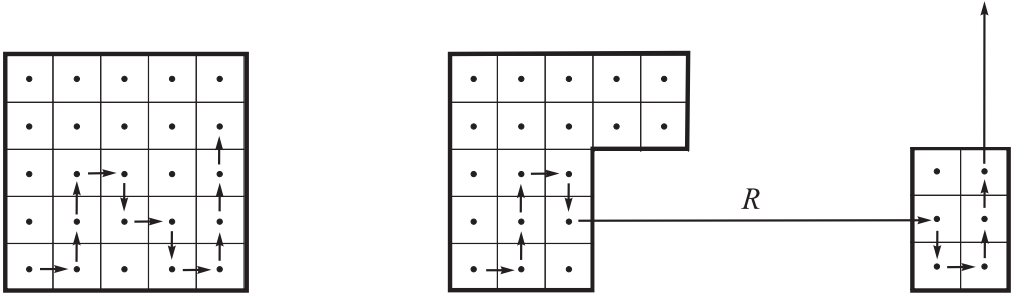}\hspace{1cm}
\caption{\small{The coarse-grained trajectory in regular (left
panel) and fractal (right panel) medium.}}
\end{figure}

However, the strongly turbulent high magnetic field is not
inherent in each cell. According to the current representation,
which was formed five decades ago, most of the space between
magnetic clouds is filled with lower quieter fields whose smooth
field lines can hold individuality at a long distance. The charged
particles of CRs move along these field lines in helical paths and
from time to time enter trap clouds, where they can stay for a
long time and ``forget'' their initial direction of motion. In
this case, a jump from one trap to another is not instantaneous as
in the above case: the particle rather intersects ``almost empty''
cells, covering large distances $R$. The distribution of $R$ in a
fractal medium involves a long power-law tail. These transitions
require the consideration of a finite velocity of the particle,
more precisely, the leading center of the particle.

\section{The bounded anomalous diffusion model}

The most important consequence of the finiteness of the velocity
of free particles is the finiteness of the spatial distribution:
the probability density beyond a sphere with the radius $vt$ and
the center at the instantaneous point source is zero in this case.
Let us refer to this process as bounded anomalous diffusion in
order to distinguish it from unbounded anomalous diffusion in
which the motion of a particle is represented as a sequence of
instantaneous jumps from one point of space to another: the
particle arrives at the latter point at the same time at which it
leaves the former point at any distance between these two points.
The delay time (in a trap) is not related to this distance and to
the motion as a whole. If the traps are removed from this model,
it becomes senseless, because the particle instantaneously flies
to infinity and leaves the system under consideration. The bounded
anomalous diffusion model involves not only the time spent in the
traps, but also the time taken for the motion of the particle and,
for this reason, is meaningful even in the absence of traps.
Finally, since the bounded anomalous diffusion propagator vanishes
beyond the sphere with the radius $vt$, all of its moments are
finite.

The effect of the finiteness of the velocity of free motion on the
walk process described above, which was investigated in
\cite{21,22,23,24,25,26}, significantly changes the continuous
time random walk model. In this case, the particle at the
observation time can be in one of two states, rest and motion. The
corresponding components of the probability density are denoted as
$\psi_0(\mathbf{r},t)$ and $\psi_1(\mathbf{r},t)$ so that the
total density is
$$
\psi(\mathbf{r},t)=\psi_0(\mathbf{r},t)+\psi_1(\mathbf{r},t).
$$
The rates of the $1\to 0$ and $0\to 1$ transitions per unit volume
near the point $\mathbf{r}$ are denoted as $F_{1\to
0}(\mathbf{r},t)$ and $F_{0\to 1}(\mathbf{r},t)$, respectively. It
is obvious that the particle passing to the state of rest at the
point $\mathbf{r}$ at the time $t-t'$ remains in this state at the
observation time $\mathbf{t}$ with the probability $Q(t') =
\int_{t'}^\infty q(t)dt$, and the particle leaving the trap at the
point $\mathbf{r} – \mathbf{r}'$ intersects a unit area at the
point $\mathbf{r}$ without an interaction with the probability
$P(\mathbf{r}') = \int _{0}^\infty
p(\mathbf{r}'+\xi\mathbf{\Omega}) d\xi, \quad
\mathbf{\Omega}=\mathbf r'/r'.$ Since this transition takes $r'/v$
seconds,
$$
\psi(\mathbf r, t)=\int \limits_0^\infty dt' Q(t') F_{1\to
0}(\mathbf r, t-t') + v^{-1} \int  d\mathbf r' P(\mathbf r')
F_{0\to 1}(\mathbf r - \mathbf r', t- r'/v), \eqno(3)
$$

The factor $v^{-1}$ appears in front of the integral because the
integral gives the flux of particles, whereas $\psi_1$ is the
concentration of these particles. The transition rates (if the
particle begins its evolution in a trap at the origin of the
coordinates at the initial time) are related as
$$
F_{1\to 0}(\mathbf r, t) = \int d\mathbf r' p(\mathbf r') F_{0\to
1}(\mathbf r - \mathbf r', t- r'/v) + \delta (\mathbf r) \delta
(t),\eqno(4)
$$
$$
F_{0\to 1}(\mathbf r, t) = \int \limits_0^\infty d\tau q(\tau)
F_{1\to 0}(\mathbf r, t-\tau),\eqno(5)
$$
The Fourier–Laplace transformation
$$
\psi(\mathbf{r},t)\mapsto \psi(\mathbf{k},\lambda)=\int
d\mathbf{r}\int\limits_0^\infty e^{i\mathbf{k}\mathbf{r}-\lambda
t}\psi(\mathbf{r},t)dt
$$
reduces the system of Eqs. (3)-(5) to the form
$$
L_v(\lambda,\mathbf{k})\psi(\mathbf{k},\lambda)\equiv[1-p(\mathbf{k},\lambda/v)q(\lambda)]\psi(\mathbf{k},\lambda)=S_v(\lambda,\mathbf{k}),
$$
where
$$
S_v(\lambda,\mathbf{k})=Q(\lambda)+
(1/v)P(\mathbf{k},\lambda/v)q(\lambda)
$$
and
$$
p(\mathbf{k}, \lambda / v) = \int p(\mathbf{r})
e^{-(\lambda/v)r}e^{i\mathbf{k}\mathbf{r}}d\mathbf{r}.
$$

The quantity $P(\mathbf{k},\lambda/v)$ is determined similarly. If
the tails of the distributions of $R$ and $T$ are of a power-law
character with exponents $\alpha\in(0,2)$ and $\beta\in(0,1]$,
respectively,
$$
\textsf{P}(R>r)\sim\frac{A}{\Gamma(1-\alpha)}r^{-\alpha},\
r\to\infty,\qquad \textsf{P}(T>t)=Q(t)\sim
\frac{B}{\Gamma(1-\beta)}t^{-\beta},\ t\to\infty.
$$
Then, using Tauberian theorems, one can show that for
$\mathbf{k}\to 0,\ \lambda\to 0$
$$
L_v(\lambda,\mathbf{k})\sim B\lambda^\beta+
\left\{%
\begin{array}{ll}
    \langle
R\rangle\langle\lambda/v-i\mathbf{k}\mathbf{\Omega}\rangle+
A\langle(\lambda/v- i\mathbf{k}\mathbf{\Omega})^\alpha\rangle, & \alpha>1, \\
    A\langle(\lambda
/v-i\mathbf{k}\mathbf{\Omega})^\alpha\rangle, & \alpha<1. \\
\end{array}%
\right.
$$
where $\mathbf{\Omega}$ is a random direction of the motion, which
is assumed to be isotropically distributed. The comparison of
different terms at $\lambda\to 0$ provides the following
conclusions.

The asymptotic expressions at $v = \infty$ for both cases have the
same form
$$
L_v(\lambda,\mathbf{k})\sim
B[\lambda^\beta+C_\infty|\mathbf{k}|^{\alpha}], \qquad
C_\infty=A|\cos(\alpha\pi/2)|/[(\alpha+1)B],
$$

This expression provides fractional differential equation (1) of
unbounded anomalous diffusion. At $v<\infty,\ \alpha=2,\ \beta=1$
(in view of the mentioned isotropy,
$\langle\mathbf{\Omega}\rangle=0$), the transport operator has the
form
$$
L_v(\lambda,\mathbf{k})\sim B\lambda^\beta+\langle
R\rangle\langle\lambda/v-i\mathbf{k}\mathbf{\Omega}\rangle+
A\langle(\lambda/v- i\mathbf{k}\mathbf{\Omega})^2\rangle
=(A/v^2)\lambda^2+(B+\langle R\rangle/v)\lambda-(A/3)k^2,
$$
corresponding to the telegraph equation
$$
\left[\frac{A}{v^2}\frac{\partial^2}{\partial
t^2}+\left(B+\frac{\langle R\rangle}{v}\right)\frac{\partial
}{\partial t}-\frac{A}{3}\triangle
\right]\psi(\mathbf{r},t)=S_v(\mathbf{r},t),
$$
which describes the bounded normal diffusion. Under the same
conditions at $\beta < 1$, the following fractional variant of the
subdiffusion telegraph equation is obtained:
$$
\left[\frac{A}{v^2}\frac{\partial^2 }{\partial t^2}+\frac{\langle
R\rangle}{v}\frac{\partial}{\partial t}+B\
_0\textsf{D}_t^\beta-\frac{A}{3}\triangle\right]
\psi(\mathbf{r},t)=S_v(\mathbf{r},t).
$$

In both bases, the form of the "deep" time asymptotic expression
for $\psi(\mathbf{r}, t)$ is independent of the velocity and leads
to the equations of bounded diffusion and subdiffusion,
respectively. The same property is observed at $\alpha<1$ when
$\beta<\alpha$. If $\alpha<\beta\leq1$ (this relation between the
parameters is used by Lagutin and Tyumentsev), the asymptotic
equation for $\psi(\mathbf{r}, t)$ has the form unusual for a
diffusion process with a pseudodifferential operator
$(\textsf{D}_t + v\mathbf{\Omega}\nabla)^\alpha$ averaged over the
directions
$$
(A/v^\alpha)\langle(\textsf{D}_t
+v\mathbf{\Omega}\nabla)^\alpha\rangle\psi(\mathbf{r},t)=S_v(\mathbf{r},t).
$$
Here, the operator
$$
\left(\textsf{D}_t+v\mathbf{\Omega}\nabla\right)^\nu
\psi(\mathbf{r},t)= \frac{1}{\Gamma(1-\nu)}
\left(\textsf{D}_t+v\mathbf{\Omega}\nabla\right)\int\limits_0^t
\frac{\psi(\mathbf{r}-v\mathbf{\Omega}(t-\tau),\tau)}{(t-\tau)^\nu}d\tau
$$
is the fractional generalization of the material derivative that
is agree with results obtained by \cite{Sok:03} for
one-dimensional L\'evy walks.

If we consider the case of homogeneous distribution
    of particles $\psi(\mathbf{r},t)\equiv\psi(\mathbf{t})$
    this operator becomes the fractional Rieman-Liouville
    derivative with respect to time:
$$
\left(\textsf{D}_t+v\mathbf{\Omega}\nabla\right)^\nu
\psi(\mathbf{r},t) \mapsto\ _0\textsf{D}^\nu_t \psi(\mathbf{t})
$$
In the stationary problem, when the function
    $\psi(\mathbf{r},t)=\psi(\mathbf{r})$ does not depend
    on time,
$$
\left(\textsf{D}_t+v\mathbf{\Omega}\nabla\right)^\nu
\psi(\mathbf{r}) \mapsto\ \left(v \mathbf{\Omega}
\nabla\right)^\nu\psi(\mathbf{r}),
$$
this operator represents fractional generalization of directional
derivative.

\begin{figure}[tbp]
\centering
\includegraphics[width=0.9\textwidth]{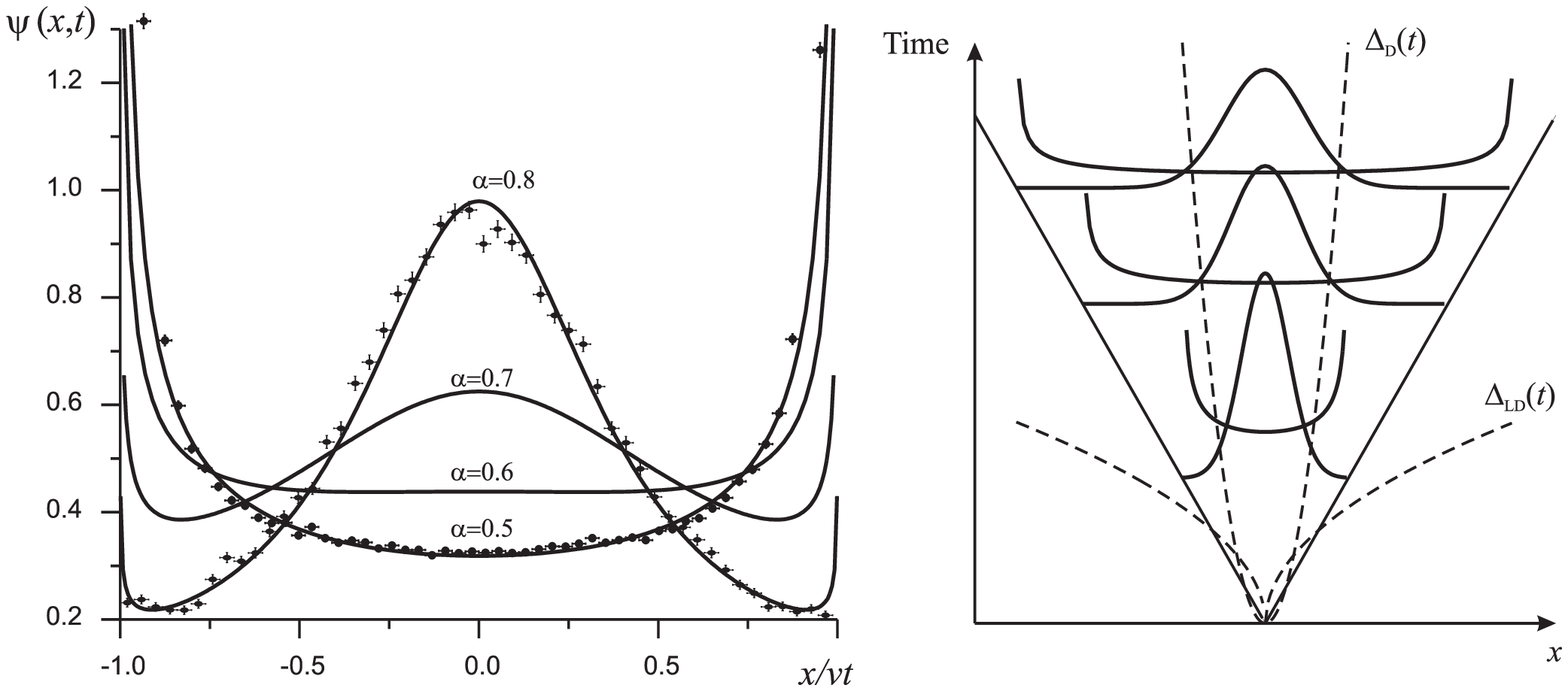}\hspace{1cm}
\caption{\small{Comparison of analytical solutions (lines) with
results of numerical Monte Carlo simulation (points) in
one-dimensional case (left panel). Evolution of the Gaussian
diffusion packet and the propagator of the bounded anomalous
diffusion model, $\alpha=0.5$ (right panel). Dashed lines
represent the time dependencies of the diffusion packet width in
both cases.}}

\centering
\includegraphics[width=0.85\textwidth]{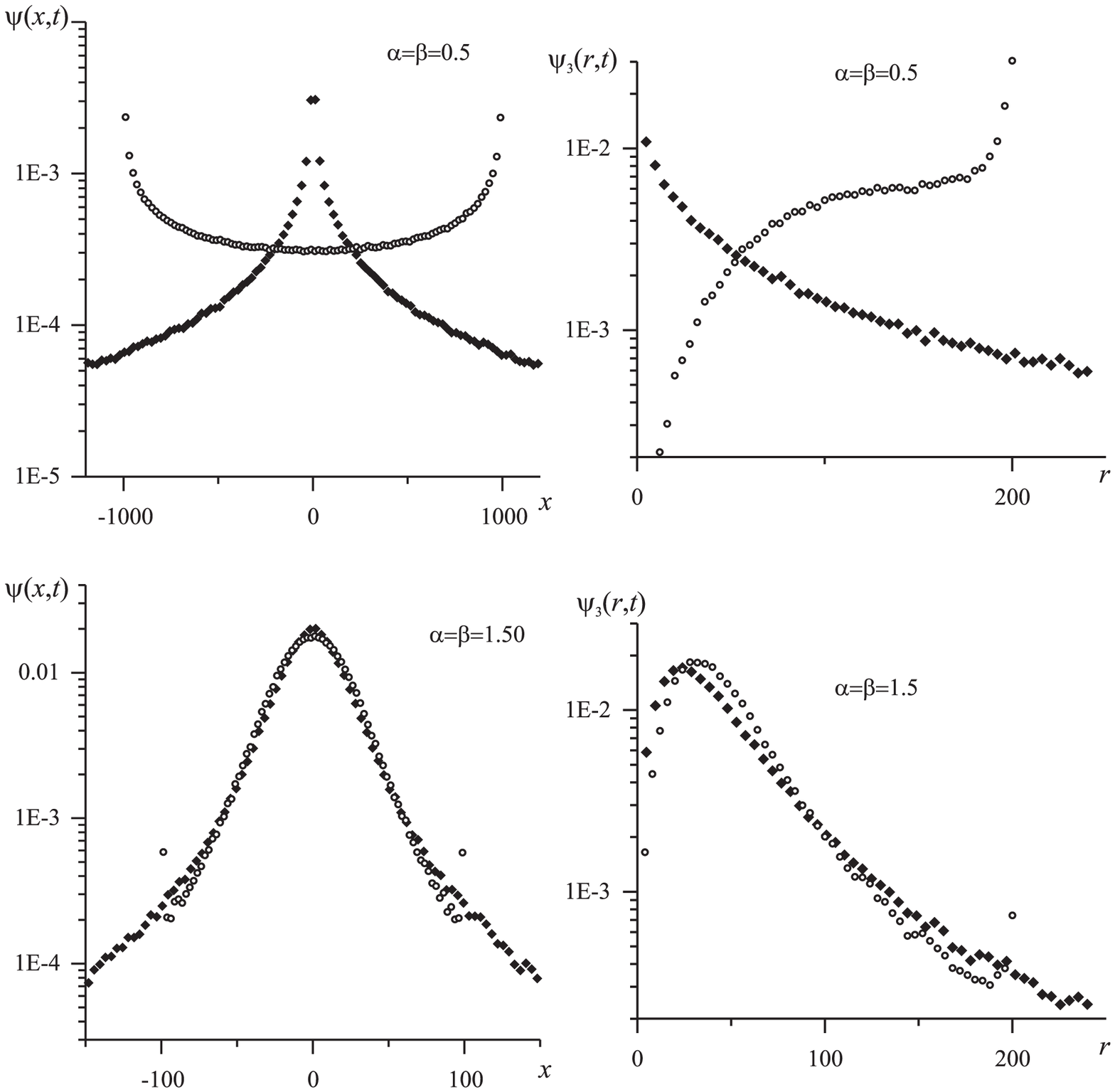}\hspace{1cm}
\caption{\small{Comparison of propagators of the bounded anomalous
diffusion (circles) and unbounded diffusion (rhombs) for
$\alpha=0.5$ and 1.5 in the 1D case (left panel) and 3D case
(right panel).}}
\end{figure}

The dependence on the velocity remains at any time, but the
dependence on â disappears. The distribution $\psi(\mathbf{r}, t)$
corresponding to this case has a specific $U$ shape in the region
bounded by the radius $vt$ beyond which it vanishes (see
\cite{25,26}).

\section{The bounded anomalous diffusion propagator}

In one-dimensional case, for $\alpha<1$, the equation of bounded
anomalous diffusion without traps takes the form:
$$
\left[\left(\frac{\partial}{\partial t}-v\frac{\partial}{\partial
x}\right)^\nu+\left(\frac{\partial}{\partial
t}+v\frac{\partial}{\partial x}\right)^\nu \right]\Phi(x,t)
=\frac{t^{-\nu}}{2\Gamma(1-\nu)}\ [\delta(x-vt)+\delta(x+vt)],
$$
Solutions of this equation can be expressed through elementary
functions (see \cite{Uch:09})
$$
\Phi(x,t)=\frac{2 \sin\pi\nu}{\pi}\frac{\left(1-x^2/v^2
t^2\right)^{\nu-1}}{(1-x/vt)^{2\nu}+(1+x/vt)^{2\nu}+2\left(1-x^2/v^2
t^2\right)^\nu \cos\pi\nu}.
$$
Solutions can also be written in terms of fractional stable
densities (see \cite{Uch:09}), that are useful for probabilistic
interpretation of these distributions.

In Fig.~2~(left panel), the analytical solutions (lines) are
compared with the results of Monte Carlo simulated random walks
with finite velocity of motion (points). In Fig.~2~(right panel),
the influence of ballistic restriction on the dynamics of
diffusion packet spreading is demonstrated schematically.

To make clear the role of correlations between path lengths and
waiting times in the model under consideration, we compare
propagators of bounded and unbounded anomalous diffusion. In the
latter model we take identical distributions for waiting times and
path lengths ($\alpha=\beta$). In the model of unbounded anomalous
diffusion path lengths and waiting times in traps are independent.
We take the model of bounded anomalous diffusion without traps,
time delay is provided by finiteness of rays propagation velocity
($v=1$). The results for the one-dimensional case are presented in
Fig.~3 (left panel), and for the 3D-model in the right panel. From
the graphs, we can see that distinction in kind between
propagators disappears only when $\alpha=\beta>2$. For values
$\alpha=\beta\leq1$, distinctions are very strong, shapes of
packets, spreading laws, behaviors near the ballistic boundaries
and near the source are essentially different. When
$1<\alpha=\beta<2$, distinctions are also sufficient despite the
fact that mean path length is finite. In one case distributions
are bounded, in another case they are unbounded. In the model of
bounded anomalous diffusion, the front near $|\mathbf{r}|=vt$
appears. The densities differ quantitatively near the source as
well. All these facts say that the results obtained on the base of
unbounded anomalous diffusion and presented in the works
\cite{13,14,15,16,27} need to be re-examined.

\section{Concluding remarks}

The work (\cite{25}, p.~1424) on anomalous diffusion ended with
the phrase: "The last property can be a reason for the conclusion
that the superdiffusion equation is inapplicable to the
description of real physical processes with the characteristic
exponents $\alpha<1$" . The cautious word ``can'' is not
accidental here: if the walk of the particle is considered in the
space of, e.g., velocities or momenta, an instantaneous jump of
the particle at large ``distance'' is allowable and even justified
in the framework of the commonly accepted model of instantaneous
collisions. The use of the model with an infinite velocity of
particles without any boundary conditions to describe the walk of
the particles in the coordinate space at $\alpha\leq\beta$ makes
both the results and conclusions doubtful (in particular, the
conclusion that ``the self-consistent description of the
experimental data on the spectra of individual groups of nuclei
and the spectrum of all of the particles is not achieve''
\cite{27} p.~601). We hope that taking into account the finiteness
of the propagation velocity of CRs in the framework of the bounded
anomalous diffusion proposed in this work will be more fruitful.

\end{document}